\documentclass[a4]{article}

\usepackage{subfigure}
\usepackage{graphicx}

\usepackage{algorithm}
\usepackage{algpseudocode}
\usepackage{color}
\usepackage{cite}
\usepackage{amssymb}
\usepackage{gensymb}
\usepackage{hyperref}

\usepackage[normalem]{ulem}

\begin{document}

\title{Time constant of the cross field demagnetization of superconducting stacks of tapes}

\author{Anang~Dadhich, Enric~Pardo,	Milan~Kapolka\\
\normalsize{Institute of Electrical Engineering, Slovak Academy of Sciences,}\\
\normalsize{Bratislava, Slovakia.}
}

\maketitle

\begin{abstract}

Stacks of REBCO tapes can trap large amounts of magnetic fields and can stay magnetized for long periods of times. This makes them an interesting option for major engineering applications such as motors, generators and magnetic bearings. When subjected to transverse alternating fields, superconducting tapes face a reduction in the trapped field, and thus it is the goal of this paper to understand the influence of all parameters in cross field demagnetization of stacks of tapes. Major parameter dependencies considered for the scope of this paper are ripple field amplitude, frequency, tape width, tape thickness (from 1 to 20 $\mu$m), and number of tapes (up to 20). This article also provides a systemic study of the relaxation time constant $\tau$, which can be used to estimate the cross-field demagnetization decay for high number of cycles. Modeling is based on the Minimum Electro-Magnetic Entropy Production method, and it is shown that the 2D model gives very accurate results for long samples when compared with 3D model. Analytical formulas for large number of cycles have been devised. The results show that when the ripple field amplitude is above the penetration field of one tape, the stack always fully demagnetizes, roughly in exponential decay. Increasing the number of tapes only increases the relaxation time. The formulas derived also hold when validated against numerical results, and can be used for quick approximation of decay constant. They also show that the cause of the decreases of cross-field demagnetization with number of tapes is the increase in the self-inductance of the magnetization currents. The trends and insights obtained for cross field demagnetization for stacks are thus very beneficial for engineers and scientists working with superconducting magnet design and applications.

\end{abstract}


\section{Introduction}

Superconducting stacks of REBCO tapes can trap upto 17.7 T field\cite{HTSTrap}. It is seen, though, that on the application of transverse field (or cross field), there is a decay in the trapped field of the sample \cite{Brandt02,campbell14SSTa,campbell17SST,baskys18SST,liangF17SST,vanderbemden07PRB,milanPre,baghdadi18ScR,fagnard16SST}, which is possible for superconducting bulks as well\cite{Kapolka18IES,srpcic19SST}. This demagnetization of superconducting tapes and stacks can have adverse effects on various electrical applications, for example, the runtimes of motors, and should be an important topic of current research. 

Superconducting motors are the next potential choice for the high energy electrical applications. The main benefits of such motors over conventional ones are reduction of size (upto 70 percent), weight, noise, and vibration. Increased efficiency is one of its another benefits and the superconducting stacks of tapes can be used in rotors of HTS motors \cite{sotelo18IEEE,climentealarcon18JCP,smara19SST,climentealarcon19JPP}. Recent advances in high temperature superconductivity and cryogenic systems have led to the use of HTS motors in various new applications, such as in aviation for future electric aircraft ( Hybrid Distributed Electric Propulsion)\cite{asumed,patel18IEEE}, generators\cite{lloberas14} (with HTS coils), marine propulsion,  and wind turbines\cite{Azo} \cite{Snitchler}. 

Some studies have been made for the cross-field demagnetization of HTS tapes \cite{Brandt02,campbell17SST}. Through Critical State Model, Brandt\cite{Brandt02} shows that for a single tape there is a decay of trapped field until an asymptotic value is reached for ripple field amplitudes below the parallel penetration field, $B_{p||}$. For ripple fields above the parallel penetration field, there is a sharp exponential decay of trapped field, resulting in full demagnetization. The cause of this behaviour is the appearance of the dynamic magneto-resistance\cite{Brandt02} \cite{jiangZ17SST}. For large ripple fields, superconducting bulks face more demagnetization as compared to the HTS stacks, and can lose upto 50 percent of magnetization after just 1 cycle of applied cross field\cite{baghdadi14ApL}. The demagnetization is also larger for rotating fields as compared to the cross fields\cite{baghdadi18ScR}.The demagnetization of the stacks is directly dependent to the ripple field amplitude and ripple field frequency, and increases linearly with the ripple field amplitude. This is due to the direct proportionality of DC electric field generated inside the superconductor to the ripple field amplitude according to Brandt and Mikitik theory \cite{Brandt02,mikitik03PRB}, atleast for high ripple fields. Demagnetization also increases with the frequency of crossed field, though the frequency dependence of the demagnetization per given number of cycles is not very drastic \cite{liangF17SST,baghdadi14ApL}. Also, for thin tapes, this magnetization decay is very slow \cite{campbell17SST}. Similarly, the relaxation decay constant is also dependent on different parameters, and it decreases with ripple field amplitude and increases with number of tapes\cite{liangF17SST,campbell17SST}. However, measurements in \cite{baskys18SST} show that for large enough ripple fields (above the parallel penetration field of one tape, according to \cite{Brandt02}) the stack fully demagnetizes after many cycles (10$^4$ or more). Since in motors for aviation the involved frequencies are atleast hundreds of Hz, 10$^4$ cycles represent to the order of 1 minute. Therefore, it is of capital importance to predict the behaviour well above 10$^4$ cycles, reaching upto millions of cycles. In order to avoid cumbersome numerical calculations, estimations could be done by extrapolating the results for a relatively low number of cycles.  

It is important to develop better computer programs which can model very thick stacks (above 20 tapes), which this paper {achieves}. We use a high mesh for our unique method, which enables us to model {the} effect of demagnetization in presence of low ripple field amplitudes {or high number of tapes}. {The use of high mesh is key to obtain accurate results of the time constant, and hence the enhanced numerical method compared to previous ones is a substantial contribution to the field.} The {dependence} of decay rate constant on various stack and field parameters that we present is also a very meaningful study for fast approximation of demagnetization rates, which can directly be used by engineers {and scientists. In addition, we develop analytical formulas for tapes, thin stacks, and thick stacks. Apart from enabling fast estimations, the physical background of these formulas provide an explanation of the causes of several observed effects, such as the increase in the time constant with the number of tapes.}     

The structure of this paper is as follows. First we derive the analytical formulas of time constant for a single tape and stack (thick and thin), and compare it with Brandt's formula\cite{Brandt02} for single tape. Then, a small introduction to modeling method and the parameters used for 2D simulations using MEMEP method is given. Later, we present our results for dependence of demagnetization and time constant of a single tape and stacks on various ripple field parameters and tape geometry. We conclude our paper by comparing the analytical formulas of time constant with numerical results. 


\section{Analytical Method}
\label{TC_Formulas}

Based on the fact that the trapped field decays exponentially for ripple fields above the parallel penetration field, the demagnetizaton decay rate constant (Time Constant) is the time taken by a superconducting stack or tape to reach 1/e{,} or around 37 percent{,} of its original magnetization after the cross field is applied, where $e$ is the Euler number. An approximated time constant formula for different cases can be derived analytically as follows.

\subsection{Time constant for a single tape}
\label{tc_singletape}

\begin{figure}[tbp]
\centering

\subfigure[][]
{\includegraphics[trim=0 0 0 0,clip,width=9 cm]{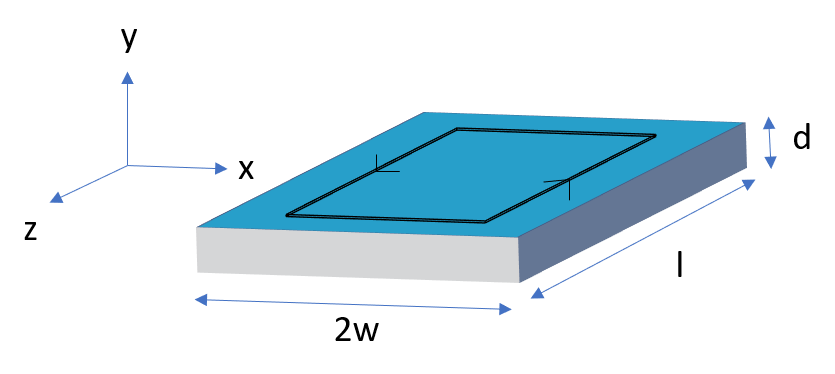}}

\subfigure[][]
{\includegraphics[trim=0 0 0 0,clip,width=7 cm]{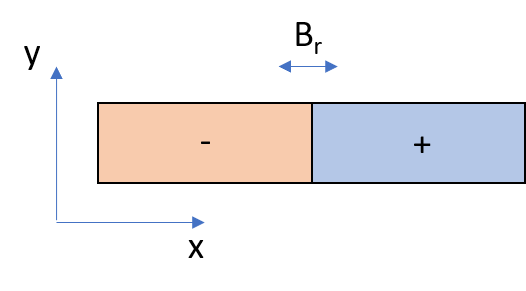}}

\caption{Qualitative sketch of the considered strip, length $l$ and width $2w$, with (a) magnetization currents, and (b) cross-section with uniform $J$ approximation.  } 
\label{geometry}
\end{figure} 

First, we assume that the tape is very long, so that the end effects are not important and, consequently, the problem can be modeled by its cross-section only. We also assume that the current density, $J$, which is equal or below the critical current density, $J_c$, is uniform in each half of the cross-section (\ref{geometry}). Although, the current density is not uniform \cite{Brandt02}, this will result in a good approximation, as we show at the end of this section and in section \ref{results}. Finally, we assume that the dynamic magneto-resistance that the ripple field creates can be predicted by the critical state model, as done in \cite{Brandt02}. 

With these assumptions, the voltage drop along the whole magnetization loop is $V=2R(I)I$, where $R(I)$ is the dynamic magneto-resistance and $I=Jwd$ (see figure \ref{geometry}). We also find that $V=-L\dot{I}$, where $L$ is the loop self-inductance and $\dot{I}=dI/dt$. Then, the differential equation for $I$ is 
\begin{equation}
2R(I)I = -L\dot{I} .
\label{diff2}
\end{equation} 
The inductance $L$ corresponds to that of a thin film with current density $+J$ on one half and $-J$ on the other half. Using that $L=\frac{1}{I^2}\int_SdSJA$, with $I$, $l$, $S$, and $A$ being the current in the circuit, the tape length, its cross section, and the vector potential respectively, we obtain

\begin{equation}
{ L = l\frac{\mu_0}{\pi}2\ln2}.
\label{ind_f}
\end{equation}
Due to the transverse AC field in a slab for $B_m$ being larger than the threshold field $B_{th}$, the dynamic magneto resistance is \cite{ogasawara76Cry,ogasawara79Cry,Brandt02,jiangZ17SST},

\begin{equation}
{ R = 2lfd\frac{1}{I_c}\bigg[B_m - B_{th}(I)\bigg] },
\label{R1}
\end{equation}
with

\begin{equation}
{ B_{th}(I) = \mu_0 J_c \frac{d}{2}\bigg(1- \frac{I}{I_c}\bigg) },
\label{bth}
\end{equation}
where $d$ and $f$ are the thickness of the tape and the frequency, respectively{, $I_c$ is the critical current relative to the main magnetization loop, and $J_c$ is the critical current density, which is assumed constant}.

Substituting $I_c$= $J_cwd$, $R$ can be written as

\begin{equation}
{ R = \mu_0lf\frac{d}{w}\bigg[\frac{B_m}{B_p} - 1 + \frac{I}{J_cwd}\bigg] },
\label{R}
\end{equation}
with $B_p$ as the penetration field of one tape in parallel applied field{,} given as
\begin{equation}
{B_p = \mu_0J_c\frac{d}{2}};
\label{Bp}
\end{equation}
where $B_m$ is the applied cross field amplitude.

Substituting $R$ from equation (\ref{R}) into (\ref{diff2}) gives the first order differential equation for the current $I$:

\begin{equation}
{ aI^2 +bI +L\dot{I} =0 },
\label{diff_eqn}
\end{equation}
where, $a$ and $b$ are constants, given as

\begin{equation}
{ a = f\mu_0\frac{2}{J_cw^2}},
\label{a1}
\end{equation}
and
 
\begin{equation}
{ b = 2f\mu_0\frac{d}{w}\bigg(\frac{B_m}{B_p} - 1 \bigg) },
\label{b1}
\end{equation}
If 
\begin{equation}
{\frac{B_m}{B_p}-1 \gg \frac{I}{J_cwd} },
\label{cond2}
\end{equation}
$R$ from equation (\ref{R}) becomes

\begin{equation}
{2\frac{R}{l} = f\mu_0\frac{d}{w}\bigg(\frac{B_m}{B_p} - 1 \bigg) },
\label{R2}
\end{equation}
which is independent of $I$. In addition, as a result of (\ref{cond2}), the term with $a$ in (\ref{diff_eqn}) can be dropped. Then, the solution of $I(t)$ is

\begin{equation}
{ I(t) = I_0 e^{-\frac{bt}{L}} },
\label{It}
\end{equation}
where, $I_0$ is the current at time $t$ = 0. From equation (\ref{It}), the time constant is

\begin{equation}
{ \tau = \frac{L}{b} }.
\label{TCaux}
\end{equation}
Thus, according to equation (\ref{b1}) and (\ref{ind_f}), the time constant $\tau$ will be

\begin{equation}
{ \frac{1}{\tau} = \frac{f\pi}{\ln 2}\frac{d}{w}\bigg(\frac{B_m}{B_p} - 1\bigg) }.
\label{TC2}
\end{equation}

Another equation for time constant is found by Brandt\cite{Brandt02} for a single tape considering $J(x)$ dependence, which is given as

\begin{equation}
{ \frac{1}{\tau} = \Lambda\frac{2\pi fd}{w}(\frac{B_m}{B_p} - 1) },
\label{TC_Brandt}
\end{equation}
where, from numerical calculations, the constant $\Lambda$ is found to be 0.6386, and $w$ is the half width of the sample. Equation (\ref{TC_Brandt}) also takes the assumption of large ripple fields or low currents into account [equation (\ref{cond2})]. 

Equations (\ref{TC2}) and (\ref{TC_Brandt}) both are very similar to each other in terms of dependencies. Equation (\ref{TC_Brandt}) is more accurate since it allows non-uniform $J$, but it is harder to derive analytically. Also, it is limited to a single tape. With our formula, given the simple assumptions, the derivation is easier and we can find more formulas regarding stacks of tapes, as can be seen below. It also enables a straightforward interpretation of the results.


\subsection{Time constant for a thin stack of tapes}

For a thin stack of very small height, the problem can be considered similar to that of a single tape. Now, we make the additional assumption that the current in the magnetization loop of each tape is the same. Then, the magnetic flux on a single tape in the stack is $\phi = nLI$, with $I$ as the current in each tape, $n$ as the number of tapes, {and} $L$ being the self inductance of one tape. Thus, the time constant is $n$ times larger than the time constant of {a} single tape.
\begin{equation}
{ \frac{1}{\tau} = \frac{f\pi}{ln2}\frac{d}{wn}\bigg(\frac{B_m}{B_p} - 1 \bigg) }.
\label{TC3}
\end{equation}
This provides a simple explanation for the linear increase of the time constant with the number of tapes observed in \cite{campbell17SST}. As we can see, this increase is simply due to the larger mutual inductance between the whole stack and the current in the magnetization loop of one of the tapes of the stack.


\subsection{Time constant for a thick stack of tapes}

{Next, we take a thick stack into account, where the height fo the whole stack, $D$, is much larger than the tape width. We also assume that the tape-to-tape separation is much smaller than the tape width.}

{For this case,} the magnetic flux crossing a tape generated by the whole stack is

\begin{equation}
{ \phi = nMI },
\label{edit}
\end{equation} 
where, $M$ is the mutual inductance between the stack and one tape, $n$ is the number of tapes, and $I$ is the current in one tape. {Assuming continuous approximation and uniform $J$,} $M$ for a stack can be found from
\begin{equation}
{ M = \frac{1}{I_{stack}I}\int_{S_{tape}}dSJA_{stack} },
\label{mut_ind}
\end{equation}
with $S_{tape}$ and $I_{stack}$ being the cross section of one tape and the current in the stack ($I_{stack}=nI$), respectively, and $J$ being the current density in one tape. For $A_{stack}$, we also assume the slab approximation, $D >> w$. This results in

\begin{equation}
{ M = l\frac{4}{3}\frac{w}{D}\mu_0 },
\label{mut_ind}
\end{equation}
where, $2w$ is the width of stack and $l$ is the length.

\begin{figure}[tbp]
\centering

{\includegraphics[trim=0 0 0 0,clip,width=9 cm]{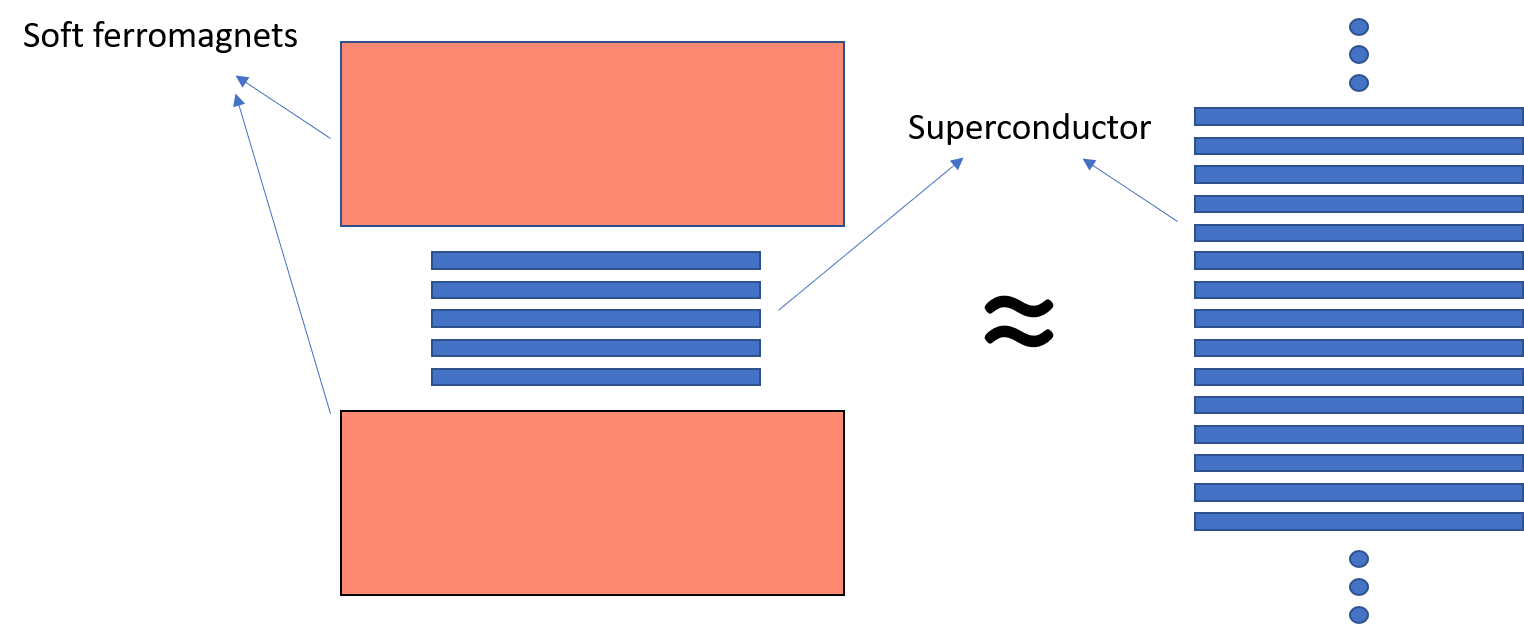}}

\caption{When a superconducting stack is kept between soft ferromagnets {(left)}, it can be assumed to be acting as an infinite stack {(right)}. } 
\label{thick_stack}
\end{figure}

From equation (\ref{edit}), the total voltage along the current loop can be given as 

\begin{equation}
{ V = \oint {\bf{E}}\cdot {\bf{dl}} = \dot{\phi} = -Mn\dot{I} },
\label{voltage}
\end{equation}
{From $V=2RI$,} the differential equation for this case and dynamic-magneto resistance $R$ is 

\begin{equation}
{ 2RI= -nM\dot{I}} .
\label{diff_eqn2}
\end{equation}
Taking the assumption of (\ref{cond2}), the solution for $I(t)$ is found to be 

\begin{equation}
{ I(t) = I_0e^{-\frac{2R}{nM}t} = I_0e^{-\frac{t}{\tau}} }.
\label{I1}
\end{equation}
Thus, from equations (\ref{mut_ind}), (\ref{R2}) and (\ref{I1}), the time constant for thick stacks of tapes is

\begin{equation}
{ \frac{1}{\tau} = \frac{3}{2}f\frac{dh}{w^2n}\bigg(\frac{B_m}{B_p} - 1 \bigg) }.
\label{TC1}
\end{equation}
For a finite stack with ferromagnetic material with high permeability present on both top and bottom sides, the system is similar to a stack with infinite number of tapes, and thus this formula can be used in this case too, which is also the case of a stack in a motor with a magnetic circuit (see figure \ref{thick_stack}) \cite{climentealarcon19JPP,smara19SST}.

\section{Modeling method}

Here, we use the Minimum Electro Magnetic Entropy Production (MEMEP) variational method \cite{pardo15SST} \cite{memep3D} to model the cross{-}field demagnetization process of superconductor stacks in 2D{, which} is a type of ${\bf{J}}$ formulation. {This method is} faster than conventional finite element methods (FEM) because the surrounding air does not need to be meshed under this model, saving many degrees of freedom\cite{roebelcomp}.

For infinitely long problems (2D), {$\bf J$} becomes a scalar, further reducing the number of degrees of freedom compared to FEM methods in the H formulation. In this sense, MEMEP has many features in common with integral methods\cite{amemiya14SST,morandi15SST,rozier19SST}. A difference is that MEMEP, as other variational methods, minimizes a functional to find the current density, and can take the multi-valued {${\bf E}({\bf{J}})$} relation of the Critical State Model into account \cite{bossavit94IEM} \cite{prigozhin97IES} \cite{memep3D}. 

Here, we use the $E-J$ power law, with power law exponent n=30, which is given as

\begin{equation}
{ {\bf{E}}({\bf{J}})= E_c \Bigg( \frac{|{\bf{J}}|}{J_c} \Bigg)^n\frac{{\bf{J}}}{|{\bf{J}}|} },
\label{Power law}
\end{equation}
where, $J_c$ is the critical current density and $E_c$ is the critical electric field. For simplicity, we assume constant $J_c$, and hence we have taken $J_c$-independent magnetic field into account in this work.

\begin{figure}[tbp]
\centering
\subfigure[][]
{\includegraphics[trim=0 0 0 0,clip,width=9 cm]{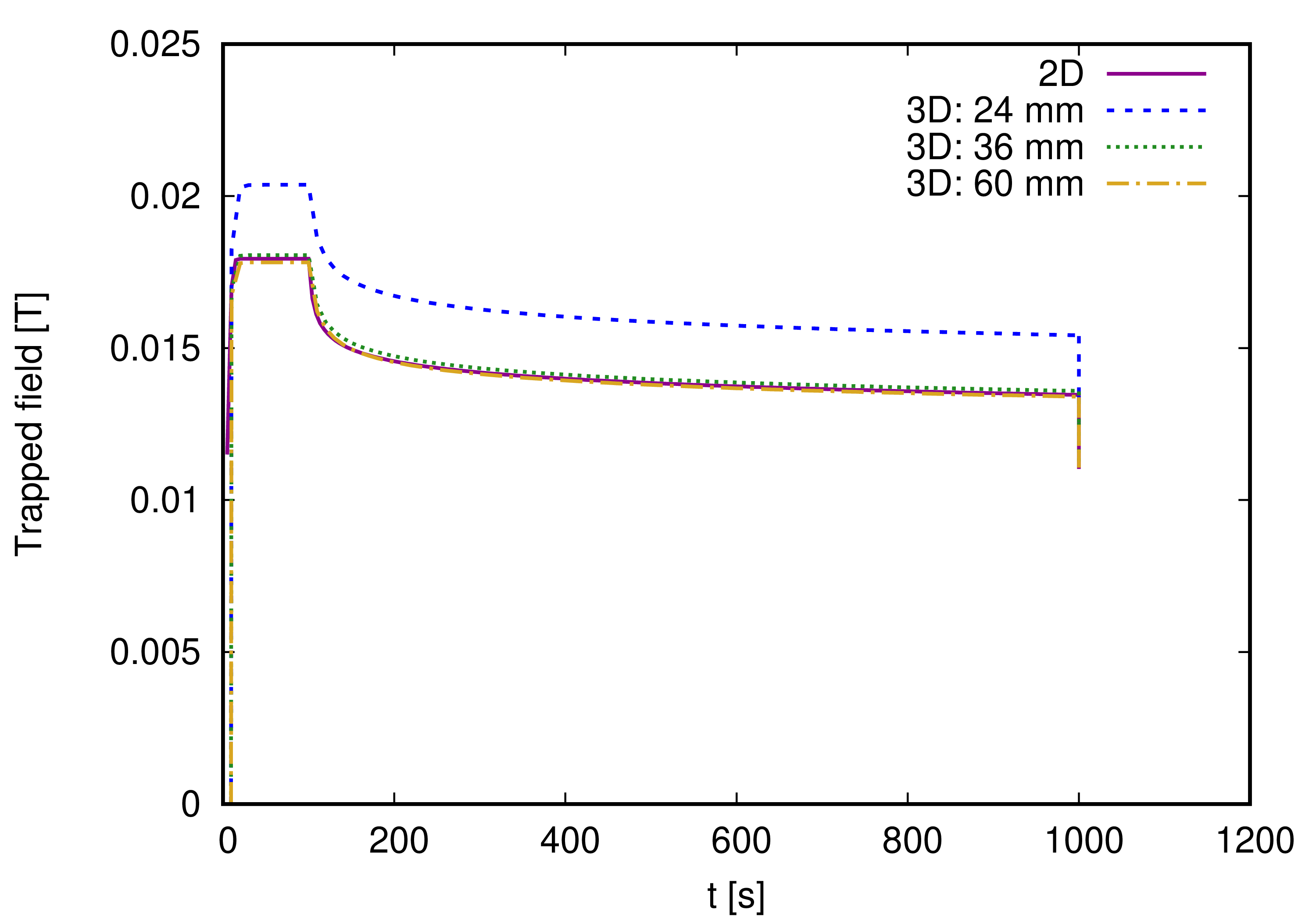}}

\subfigure[][]
{\includegraphics[trim=0 0 0 0,clip,width=9 cm]{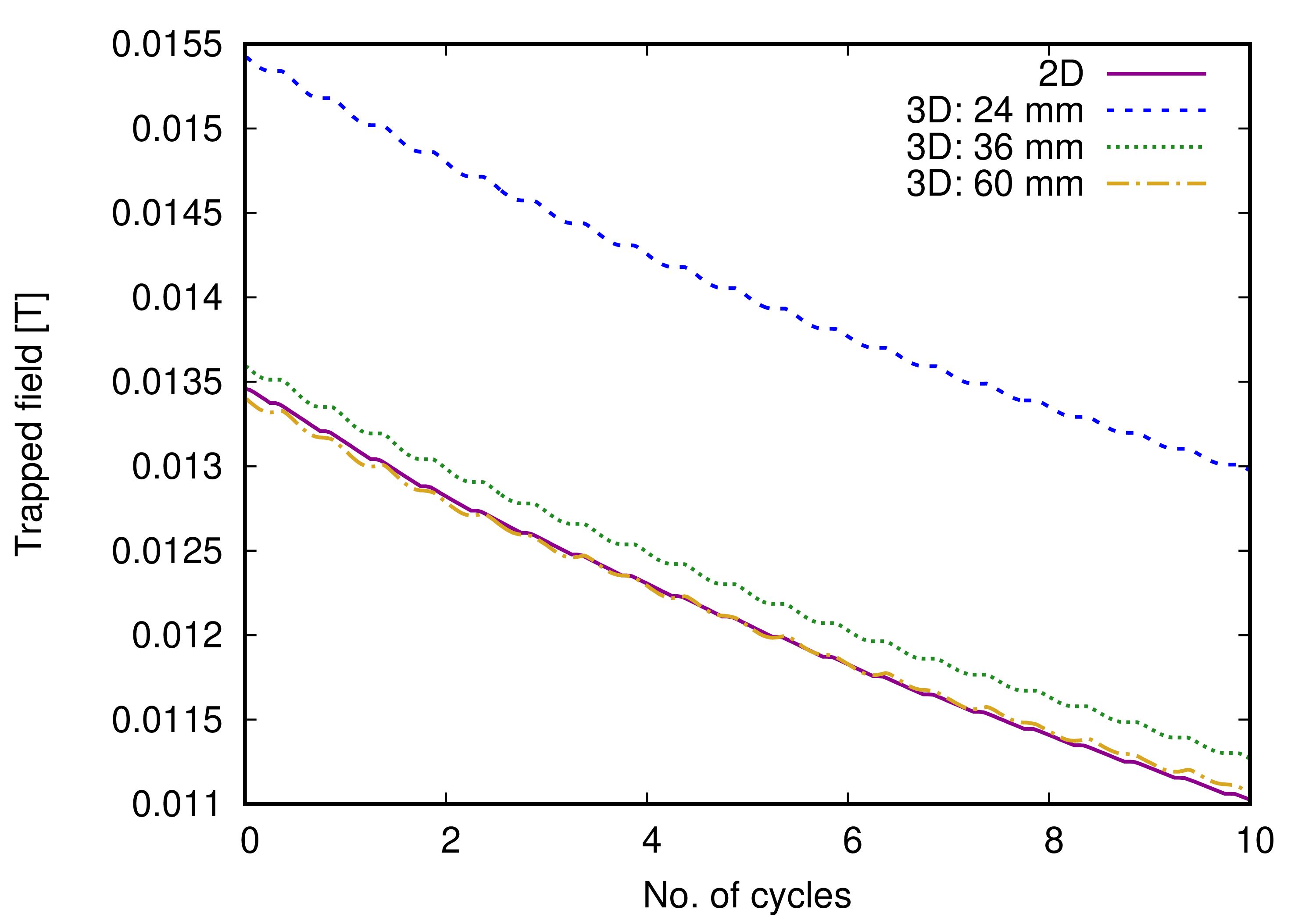}}

\caption{{The }comparison between 3D and 2D models {shows that the 2D model is realistic for tape lengths 3 times or more of the tape width. Graphs are} for (a) the whole trapped field behaviour, and (b) trapped field behaviour during cross field demagnetization process. Different lengths of tape {are} used for {the} 3D model{, while the} 2D model {assumes} infinite length. The width of the tape is 12 mm for both 2D and 3D {models}.} 
\label{3Dvs2D}
\end{figure}

{Here, we compare the} 2D model {from this work} and {the} 3D model {from \cite{memep3D,kapolka20SST}} for benchmarking purposes. The width used for the tape is 12 mm for both 3D and 2D model. The 3D model uses different lengths (24 mm, 36 mm, 60 mm), whereas the 2D model uses infinite length for the tape. From figure \ref{3Dvs2D}(a) and (b), it can be seen that for the tape lengths three times or more than the width of the tape, the trapped field values are practically the same for both 2D and 3D models. Hence, 2D model can be used for long tape samples.


\section {Modeling Configuration}

The magnetization behavior of a single tape and stacks of tapes is analyzed in the next section, under various conditions. By default, the tape thickness, width, and separation between tapes are considered to be 2 $\mu$m, 12 mm, and 60 $\mu$m respectively. For the dependence on the thickness we take thicknesses between 1 $\mu$m to 20 $\mu$m into account. The sample is initially magnetized by Field Cool process for 100 seconds under 300 mT applied field amplitude. Then{,} the sample is let to relax for 900 seconds. Later{,} a cross field is applied to the sample. Unless specified, values are 200 mT amplitude, 500Hz frequency, and maximum number of cycles 30, although some calculations reach up to 250 cycles. The critical current density $J_c$ for the sample is considered to be 1.36 $\times$ 10$^{10}$ A/m$^2$, and the trapped field is observed at 1 mm distance from the center of the surface of the stack. The {standard} mesh consists of 24 elements in thickness and 40 elements in width{, although for} some cases the mesh {reaches up to 200 elements in the tape thickness} (figure \ref{tc_all} (a) and (d)). The parallel penetration field of the tape, according to the slab model \cite{bean62PRL} \cite{zeldov94PRB}, is

\begin{equation}
{ B_{p||}= \mu_0 \frac{J_cd}{2} },
\label{Pen field}
\end{equation}
being 17 mT for $d =$ 2 $\mu$m and our chosen $J_c$. The simulations are performed on a 64 bit Linux operating system based computer with i7-7700 processor, having 3.60GHz x 8 logical cores and 16 GB RAM. With this machine, the computation times for a single 2 $\mu$m tape, using high mesh (200 x 10 elements), is around 24 hours for 30 cycles and 20 time steps per cycle. For a 10-tape stack using same parameters, the results take up to 1-1.5 weeks, depending on the ripple field amplitude.  


\section{{Modelling} Results and Discussion}
\label{results}

\begin{figure}[tbp]
\centering
\subfigure[][]
{\includegraphics[trim=0 0 0 0,clip,width=10 cm]{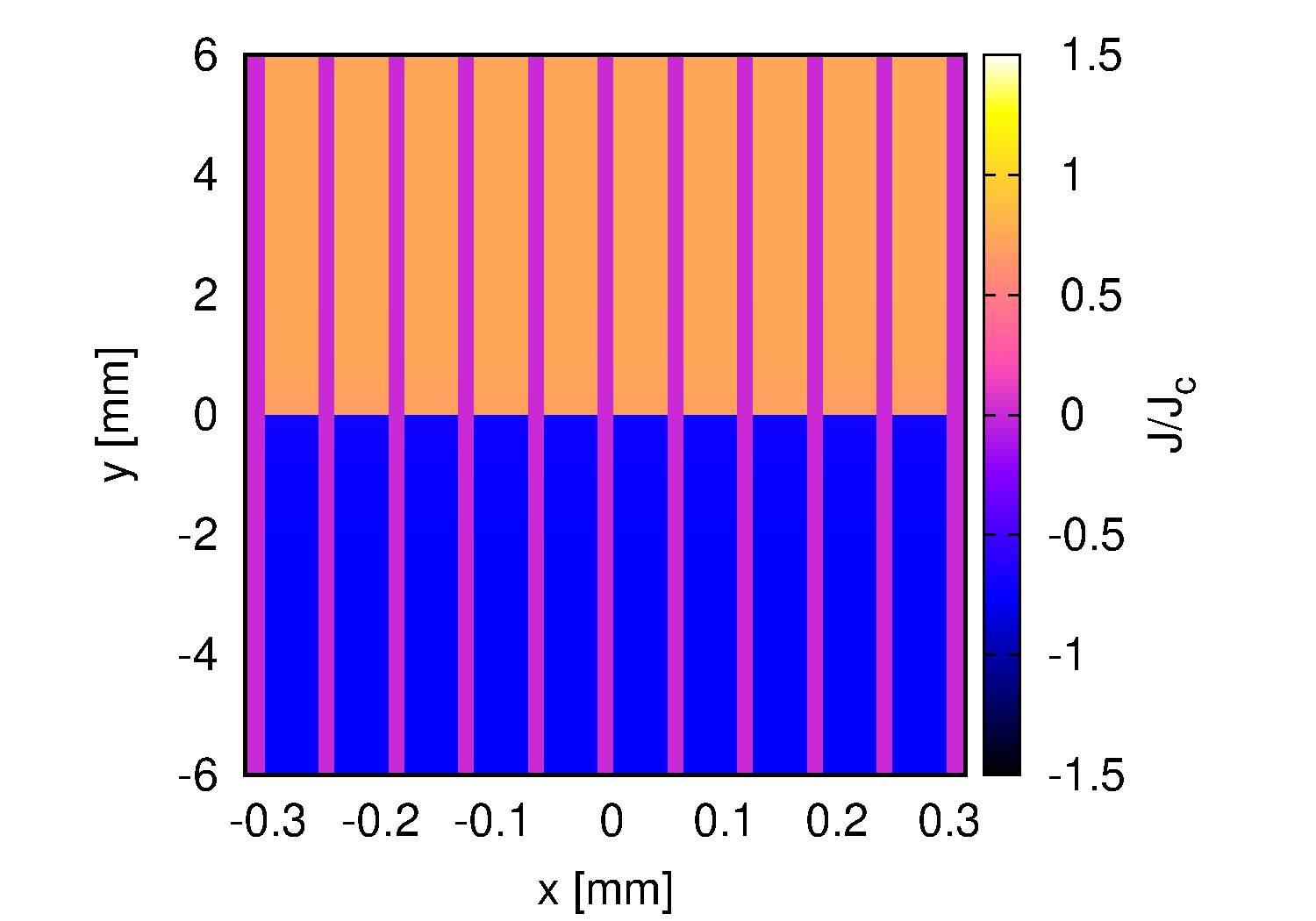}}

\subfigure[][]
{\includegraphics[trim=0 0 0 0,clip,width=10 cm]{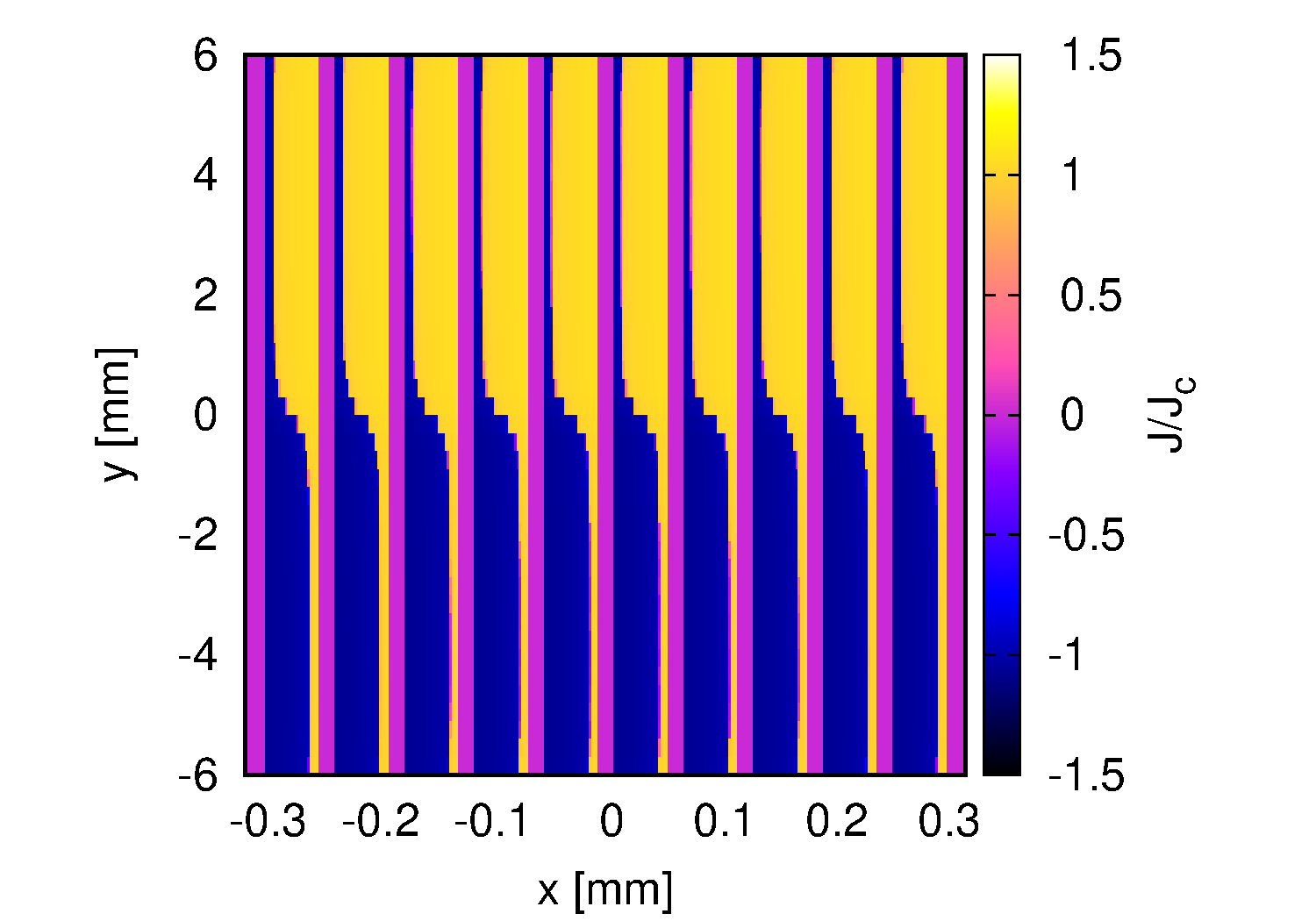}}

\caption{Current density profiles for a stack of 10 tapes, with single tape thickness 2 $\mu$m, at (a) end of relaxation and magnetization (1000 seconds), and (b) end of Cross Field Demagnetization (200 mT, 30 cycles). Legend normalized by critical current density $J_c$ = 1.36$\cdot$10$^{10}$ A/m$^2$. Tape thickness expanded in plot for better visibility.} 
\label{curpro1}
\end{figure}

Figure \ref{curpro1} shows the demagnetization behavior of the current density in a stack of 10 tapes. The stack is fully saturated by the end of magnetization and relaxation period. It is observed that there is significant demagnetization in the stack after application of 30 cycles of cross ripple field. 

\begin{figure}[tbp]
\centering

\subfigure[][]
{\includegraphics[trim=0 0 0 0,clip,width=8.5 cm]{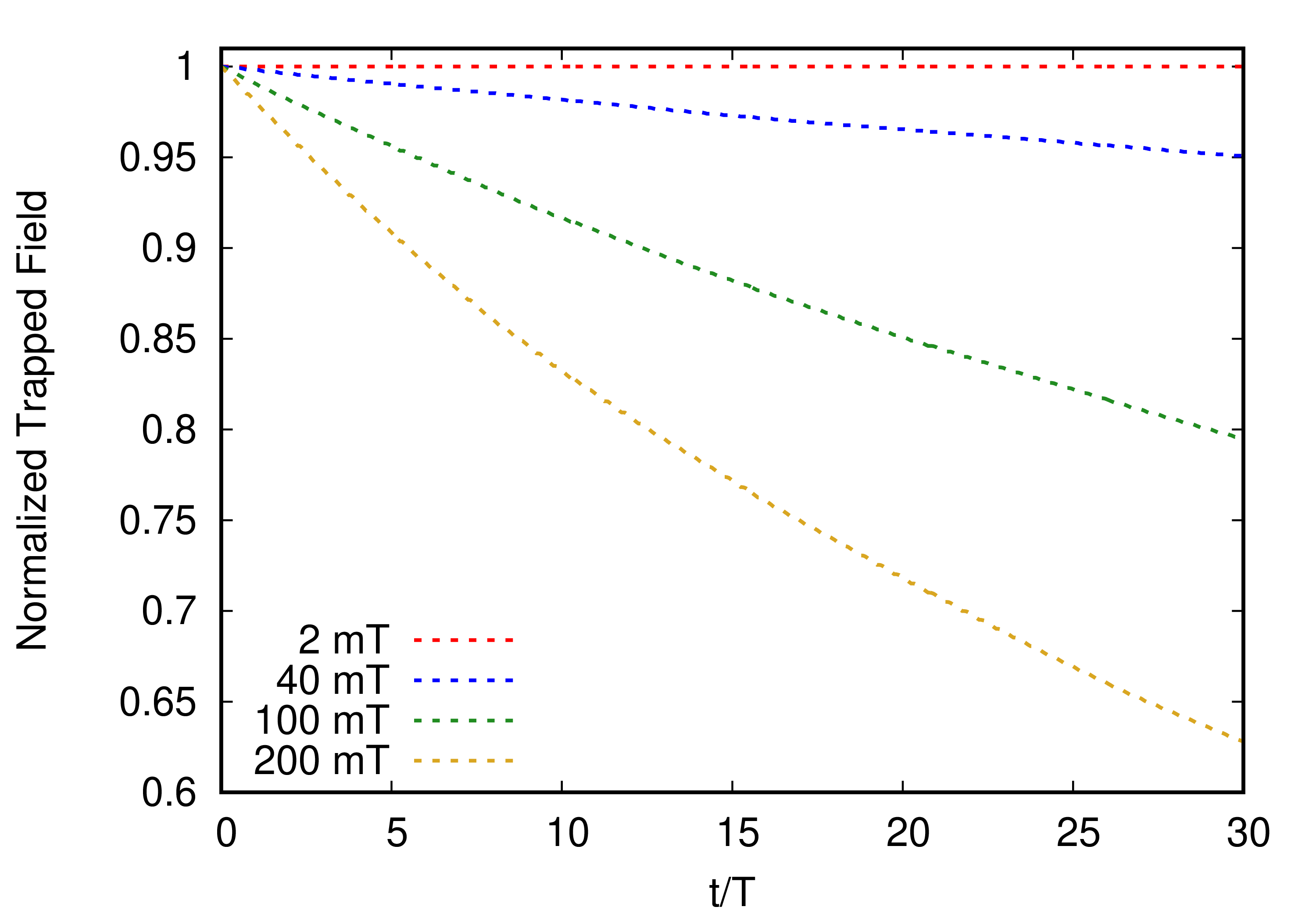}}

\subfigure[][]
{\includegraphics[trim=0 0 0 0,clip,width=8.5 cm]{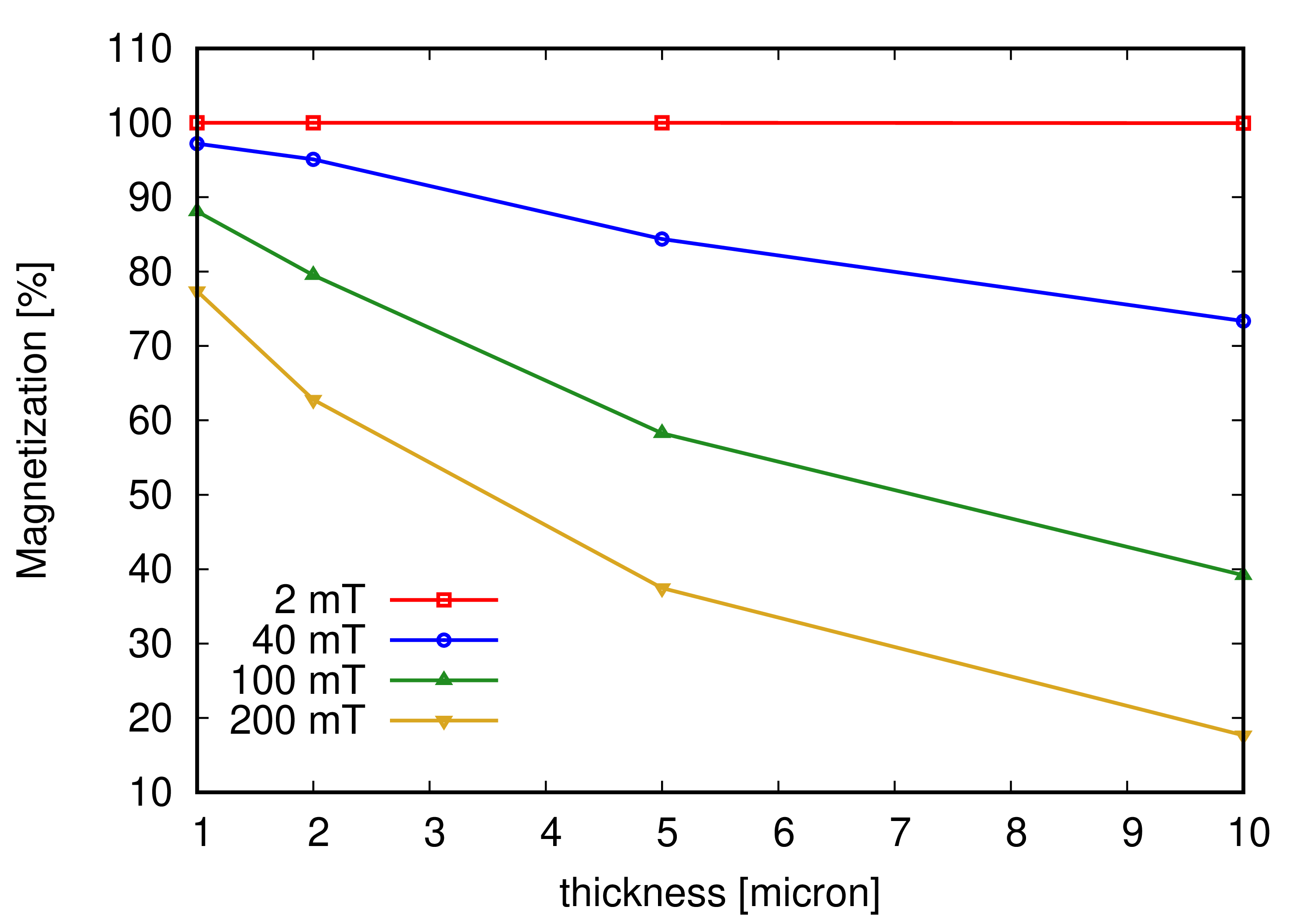}}

\caption{ (a) Trapped field curves for {one} 2 $\mu$m {thick} tape, and (b) thickness dependence of tape at different ripple fields during cross field demagnetization, constant $J_cd$, 30 cycles } 
\label{demag_thickness}
\end{figure}

The dependence of the demagnetization of a single tape on thickness and ripple field amplitude for constant sheet critical current density ($J_cd$) is shown in Figure \ref{demag_thickness}. The demagnetization below penetration field of the tape ( 17 mT) is negligible, but the trapped {field} decay is higher for high ripple fields [Figure \ref{demag_thickness}(a)]. It is also seen from Figure \ref{demag_thickness}(b) that, for constant sheet critical current density, $J_cd$, the magnetization decreases with increase in thickness {being this behavior} more evident for higher ripple field amplitudes. Thus, we see that the real thickness is very important for modeling superconductors numerically, and {that} artificial thickness should not be used in the case of cross field demagnetization studies.   

\begin{figure}[tbp]
\centering
{\includegraphics[trim=0 0 0 0,clip,width=8.5 cm]{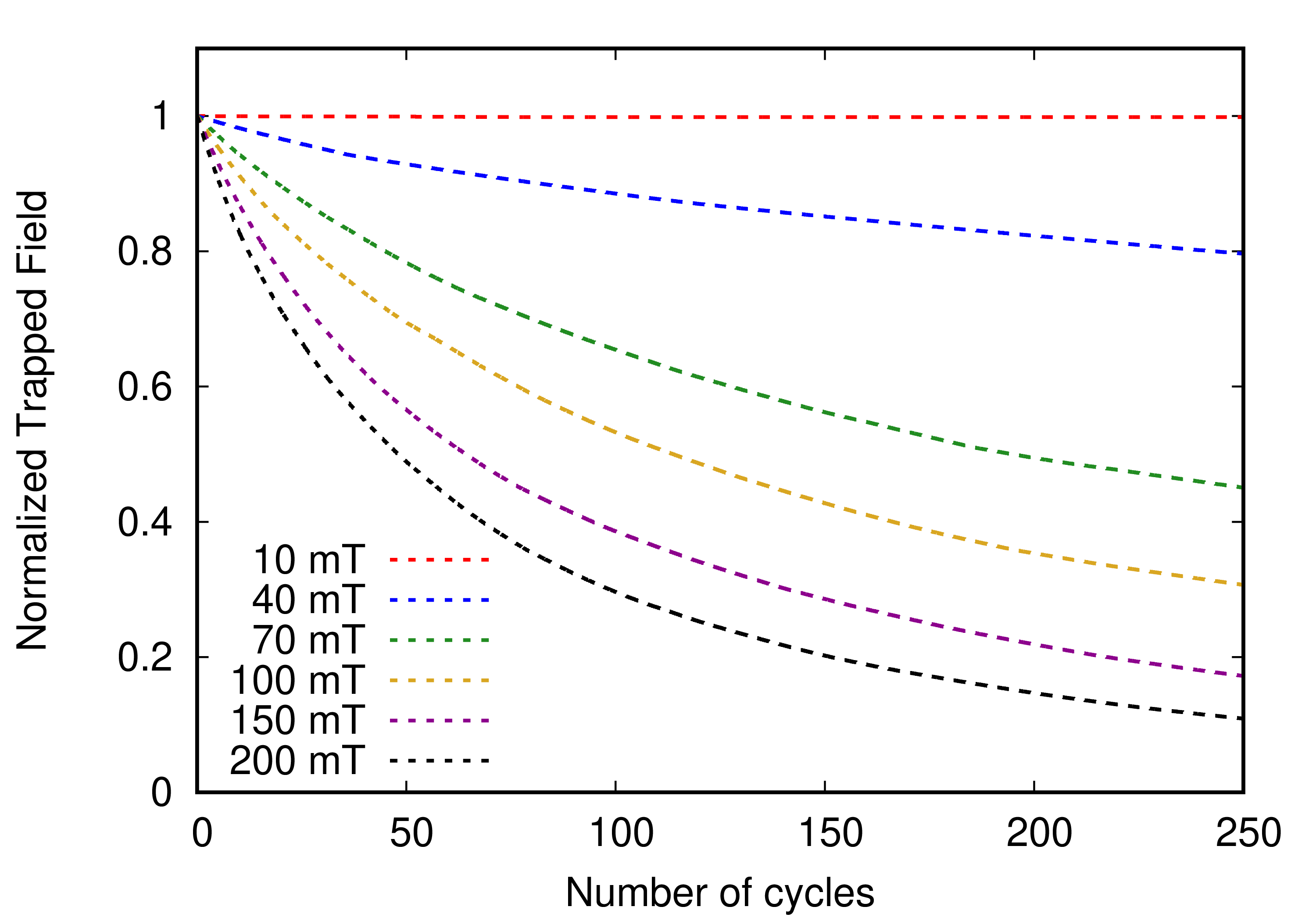}}
\caption{ Trapped field dependence with change in ripple field amplitude for tape thickness 2 $\mu$m, constant $J_c$=1.36$\cdot$10$^{10}$ A/m$^2$ for 250 cycles } 
\label{trap_B}
\end{figure}

This dependence on ripple field amplitude can also be seen for the constant $J_c$ case in Figure \ref{trap_B}, for high number of cycles. It is observed that the trapped field decay is in exponential form above the parallel penetration field of the 2 $\mu$m tape, in accordance to \cite{Brandt02}. This is a very important feature, since from the initial trapped field value, $B_{t0}$, we can extrapolate the trapped field curve as

\begin{equation}
{ B_t = B_{t0} e^{-t/\tau}},
\label{TC_num}
\end{equation}
where $\tau$ is the time constant, as discussed in section \ref{TC_Formulas}, and $B_t$ is {the} trapped field at any given time $t$. However, we should keep in mind that for ripple field amplitudes below the penetration field, the decay is no longer exponential, reaching an asymptotic value for very high number of cycles \cite{Brandt02}. Then the calculated time constants for ripple fields below the penetration fields should be regarded as pessimistic.

\begin{figure}[tbp]
\centering
\subfigure[][]
{\includegraphics[trim=0 0 0 0,clip,width=8.5 cm]{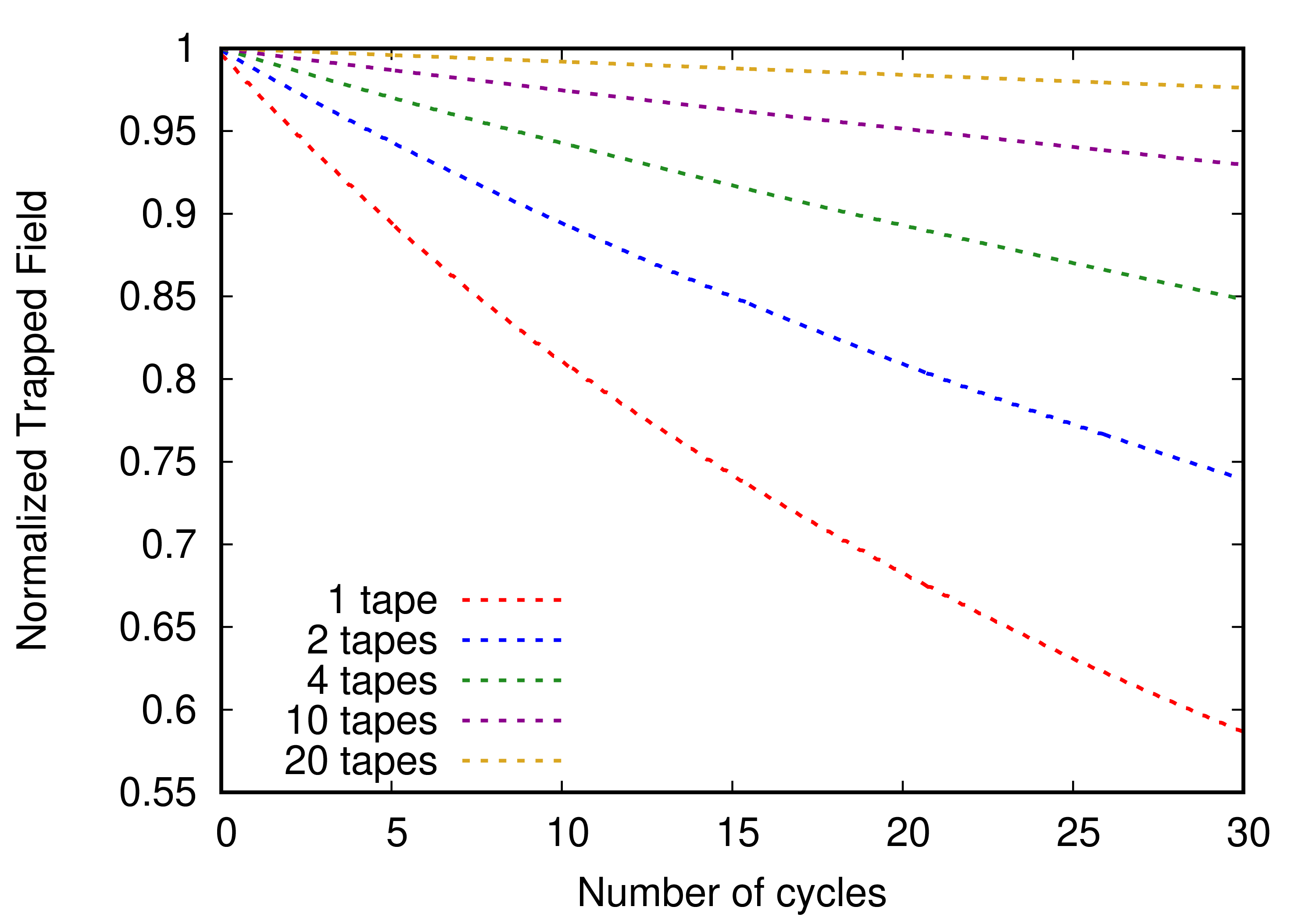}}

\subfigure[][]
{\includegraphics[trim=0 0 0 0,clip,width=8.5 cm]{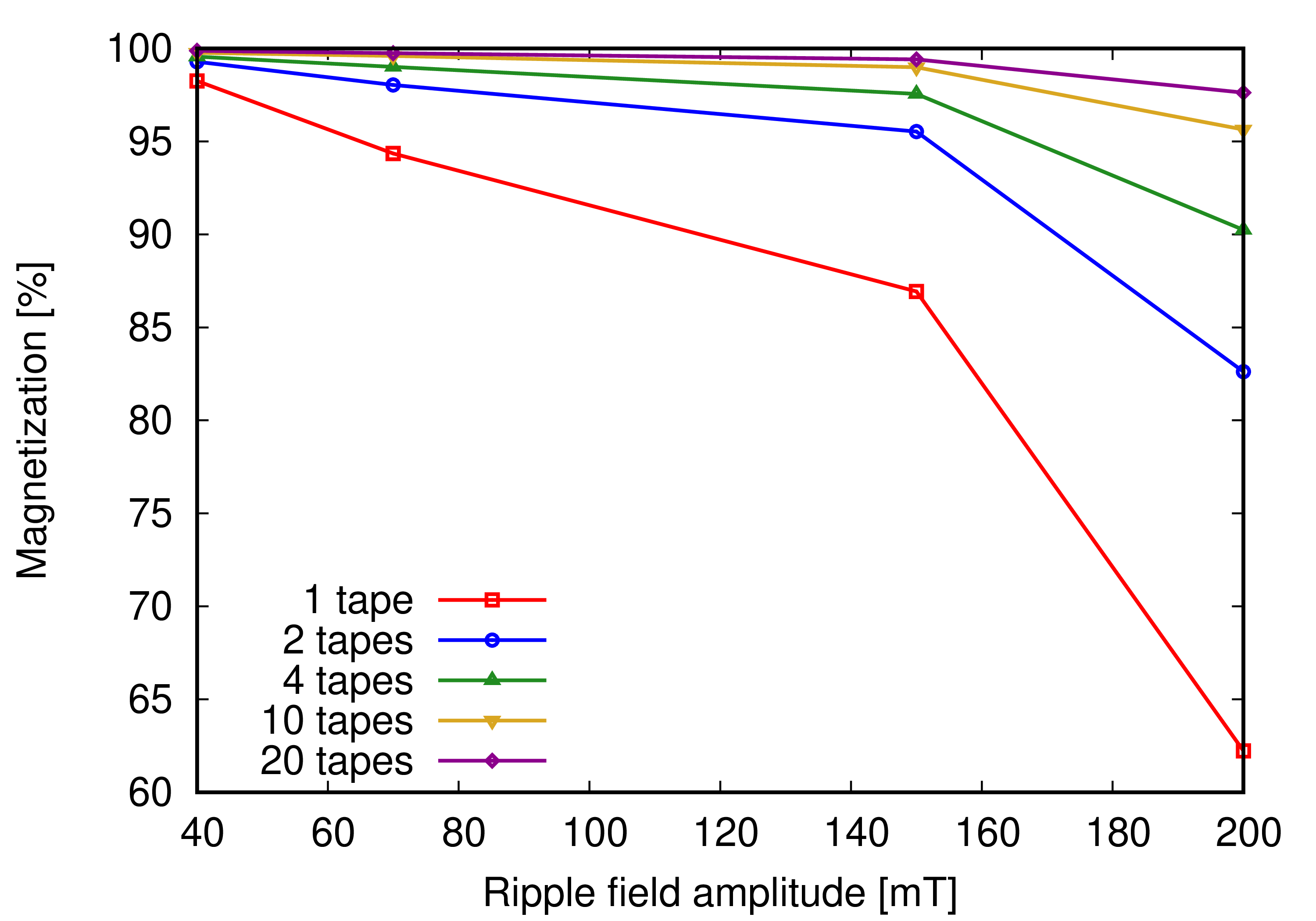}}
\caption{ Number of tape dependence on (a) Trapped field profile for 200 mT ripple field amplitude, and (b) Demagnetization for different ripple field amplitudes. Tape thickness 2 $\mu$m, 30 cycles, $J_c$=1.36$\cdot$10$^{10}$ A/m$^2$ } 
\label{demag_tape}
\end{figure}

The demagnetization also depends on the number of tapes. From Figure \ref{demag_tape}, it is observed that the demagnetization decreases with number of tapes, with the magnetization of {a} 20-tape stack reducing by only about 3 percent after 30 cycles at 200 mT amplitude. For lower field amplitudes, this decay is even lower for the 20-tape stack. The reason of the decrease in demagnetization {rate} with the number of tapes in the stack is the increase in the self inductance of the magnetization currents{, as seen in section \ref{TC_Formulas}}.

\begin{figure}
\centering
\subfigure[]{
\label{fig:first}
\includegraphics[width=6.0 cm]{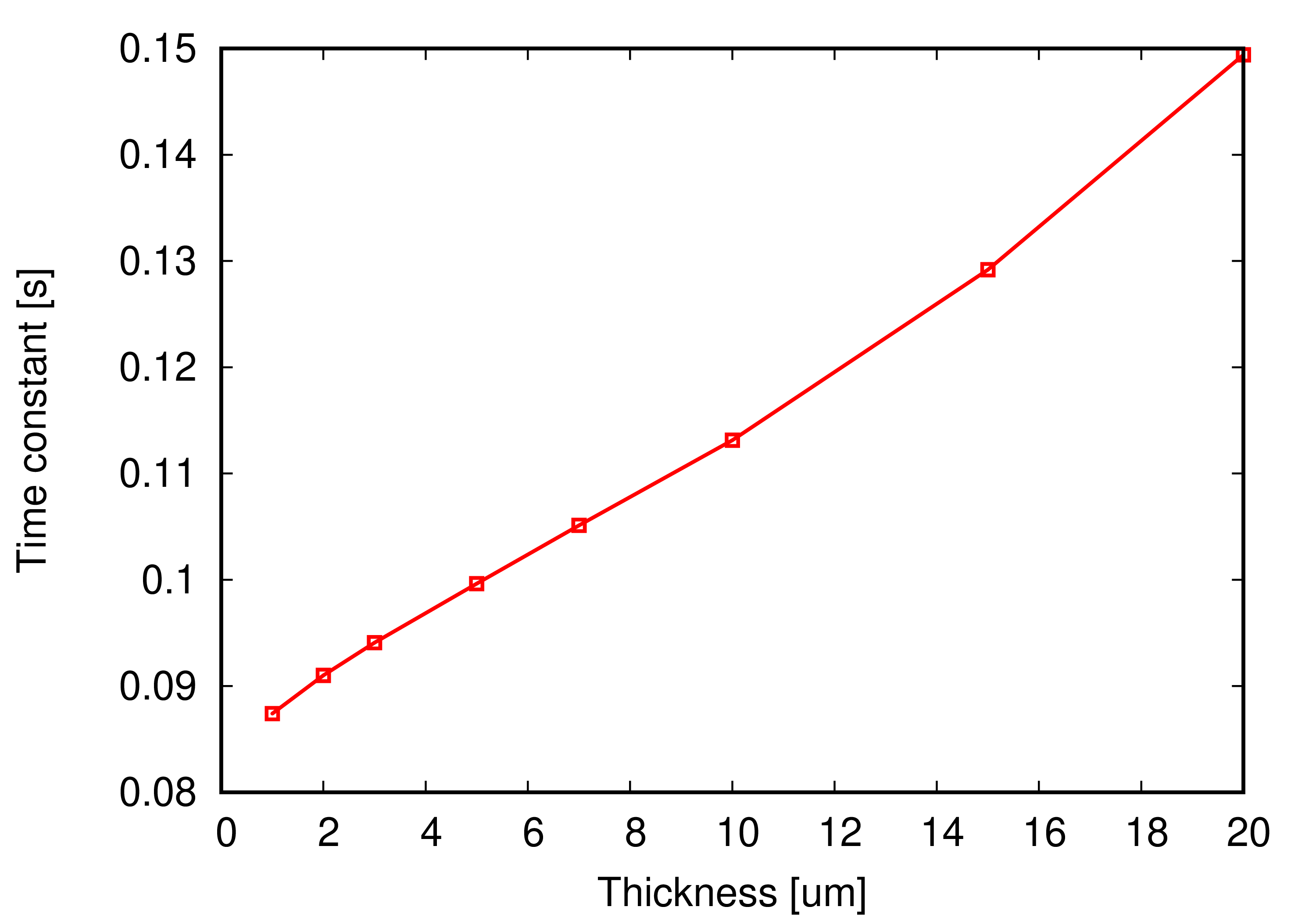}}
\qquad
\subfigure[]{
\label{fig:second}
\includegraphics[width=6.0 cm]{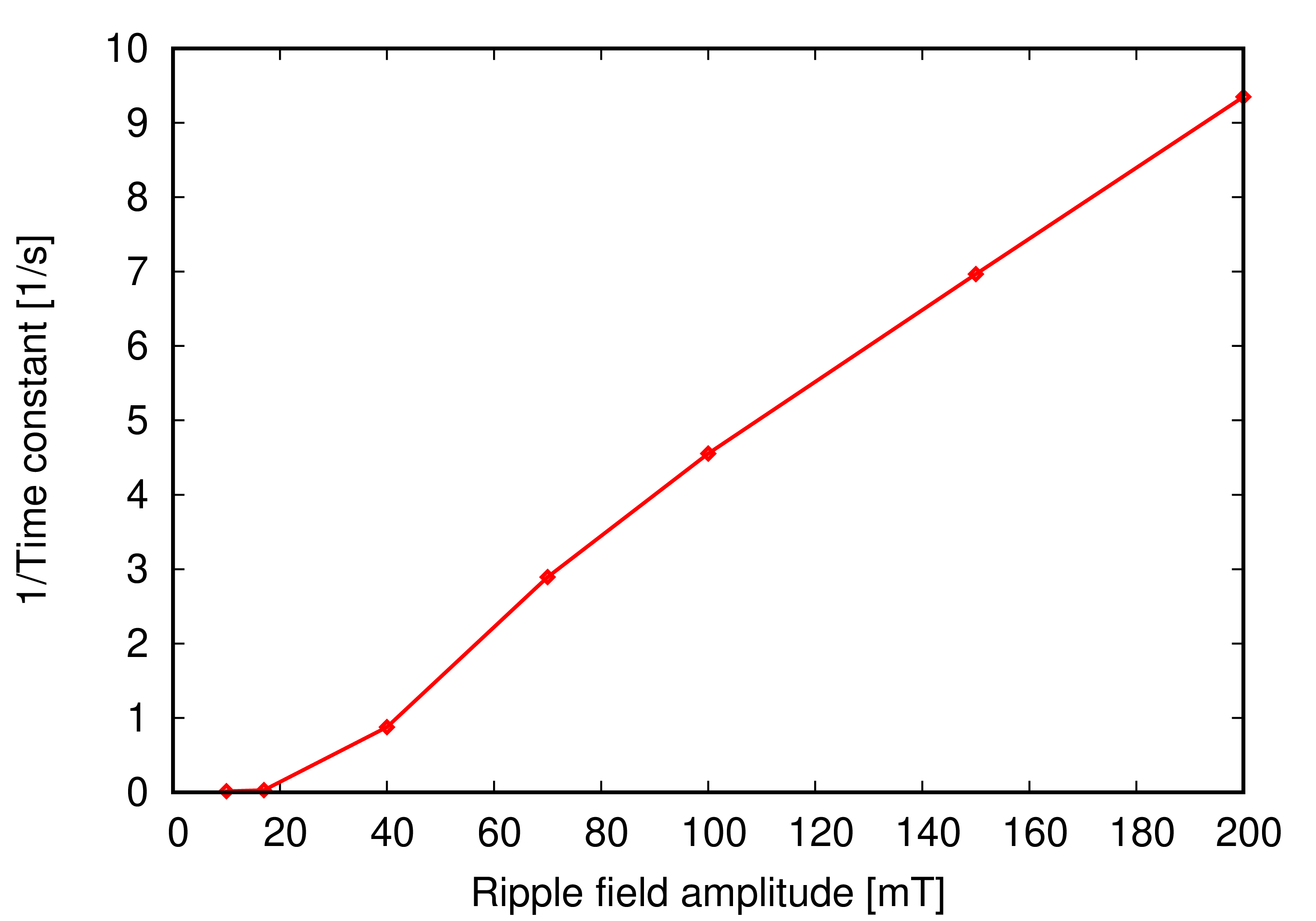}}
\subfigure[]{
{\includegraphics[trim=0 0 0 0,clip,width=6.0 cm]{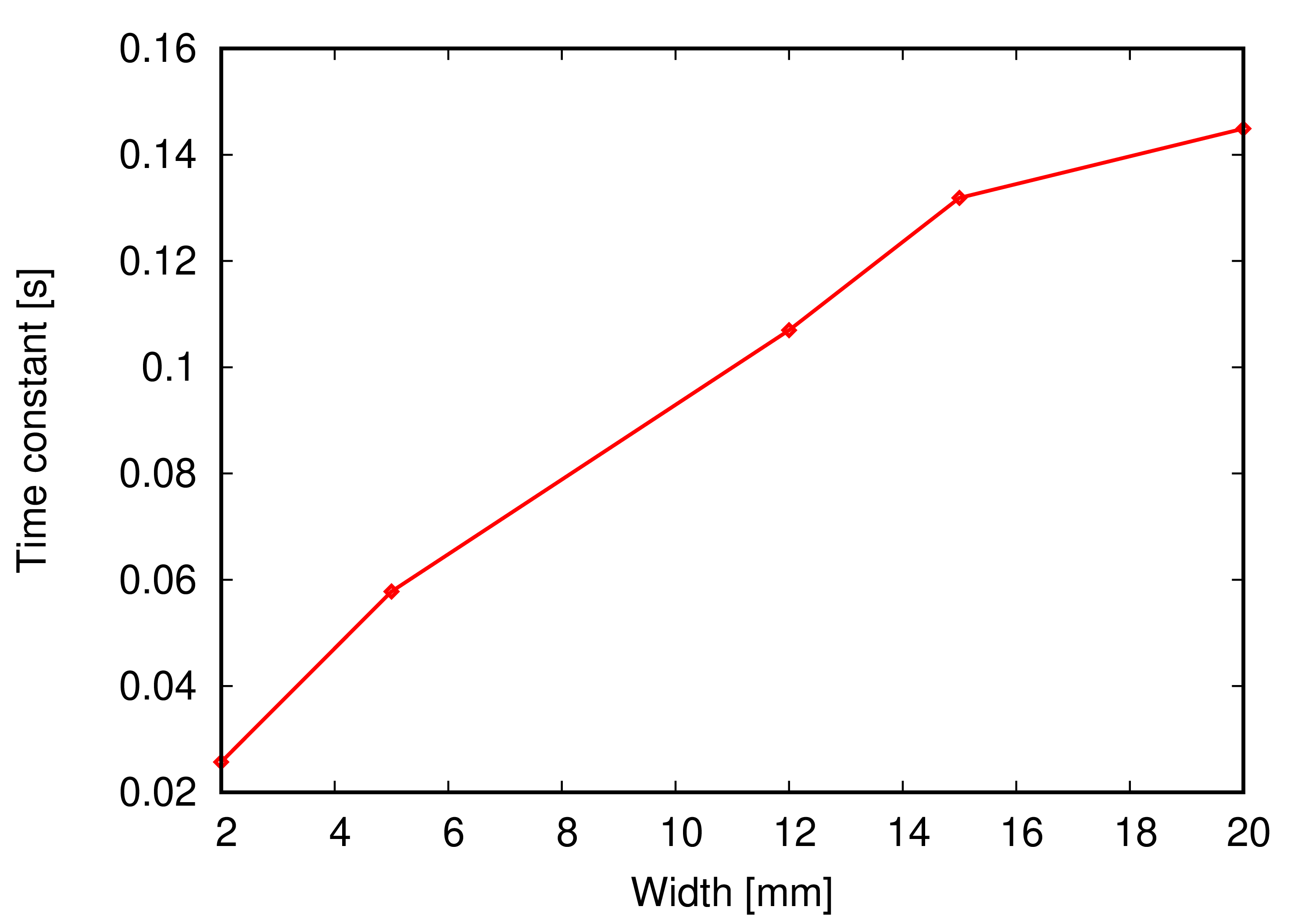}}}
\subfigure[]{
{\includegraphics[trim=0 0 0 0,clip,width=6.0 cm]{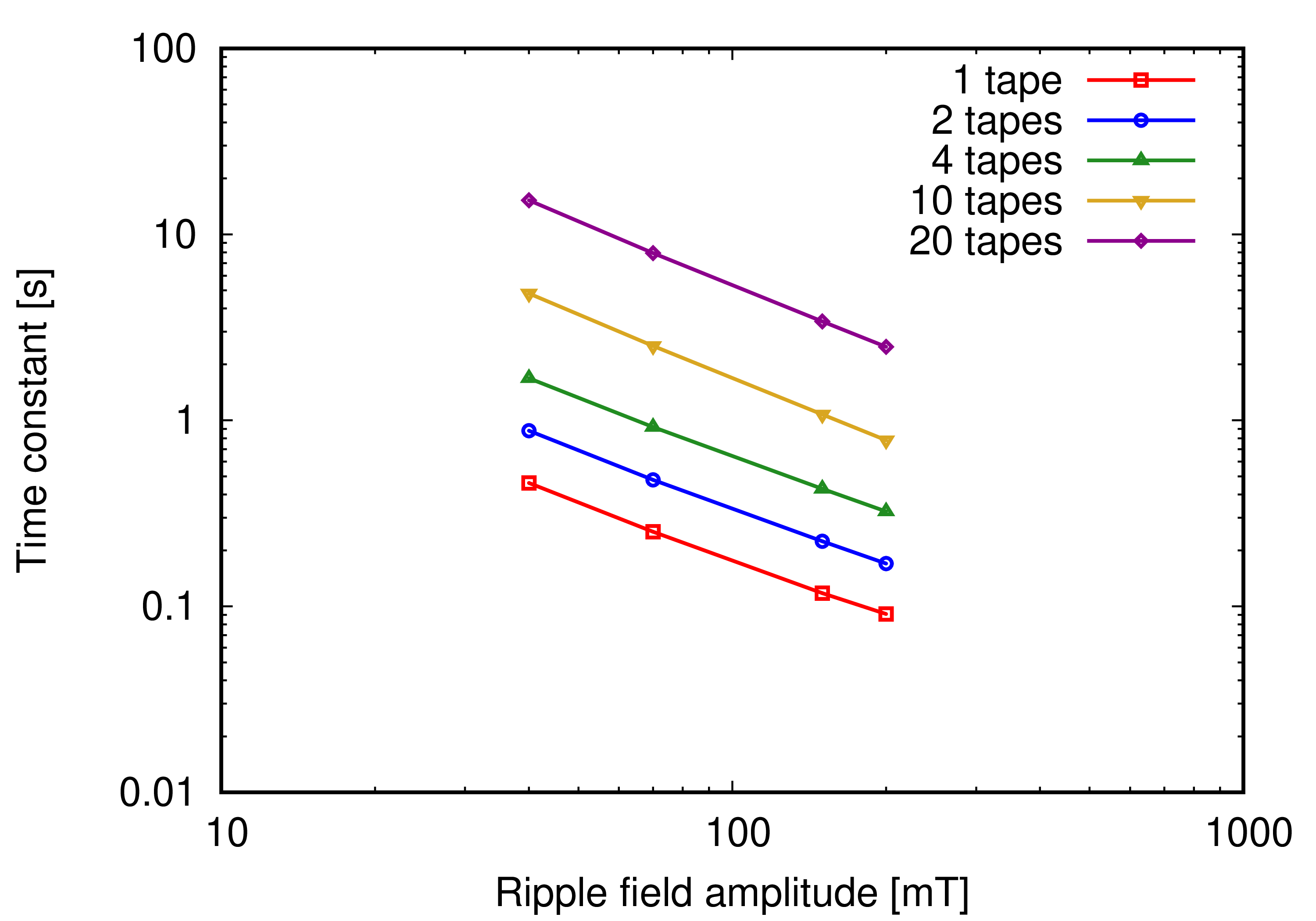}}}
\caption{ Time Constant dependence with change in (a) thickness of tape, (b) ripple field amplitude, (c) width of tape, and (d) number of tapes. Constant $J_c$=1.36$\cdot$10$^{10}$ A/m$^2$, 30 cycles. Mesh used for (a) and (d) is 200 x 10 elements, with 200 in thickness and 10 in width. } 
\label{tc_all}
\end{figure}

Given the exponential behavior of the trapped field curves, the time constant analysis can be done for a single tape and a stack of tapes for constant $J_c$. From Figure \ref{tc_all} (a), it can be seen that the time constant is higher for thicker tapes, and increases with thickness. This is in contrast to the previous result in Figure \ref{demag_thickness} for constant $J_cd$ case where {the} time constant {decreases} with thickness. For constant $J_c$ case, this {improvement} is due to the increase of penetration field (from 8.5 mT to 170 mT), which in turn reduces the dynamic magneto resistance, and hence the increase in time constant is found. Then, recent advances in increasing the superconductor thickness in REBCO for nearly the same $J_c$ has beneficial consequences regarding cross field demagnetization.

Similarly, the time constant decreases with ripple field amplitude due to dynamic magneto-resistance, and hence $1/\tau$ shows a linear behavior for amplitudes over penetration field for a single tape. The time constant is also width dependent, increasing with the width of tape. {The cause is now the reduction in the dynamic magneto-resistance associated to the main magnetizing loop [see equation (\ref{bth})].} Note that the tape self-inductance in equation (\ref{ind_f}) is independent {on} tape width.

For the stack of more than one tape, the time constant is directly dependent on number of tapes, and goes higher with more tapes in a stack, as can be seen in Figure \ref{tc_all}(d). At higher field amplitudes, there is still some decay in {the} 20-tape stack, but the time constant values are still much higher as compared to that of a single tape.

\begin{figure}[tbp]
\centering
\subfigure[][]
{\includegraphics[trim=0 0 0 0,clip,width=8.5 cm]{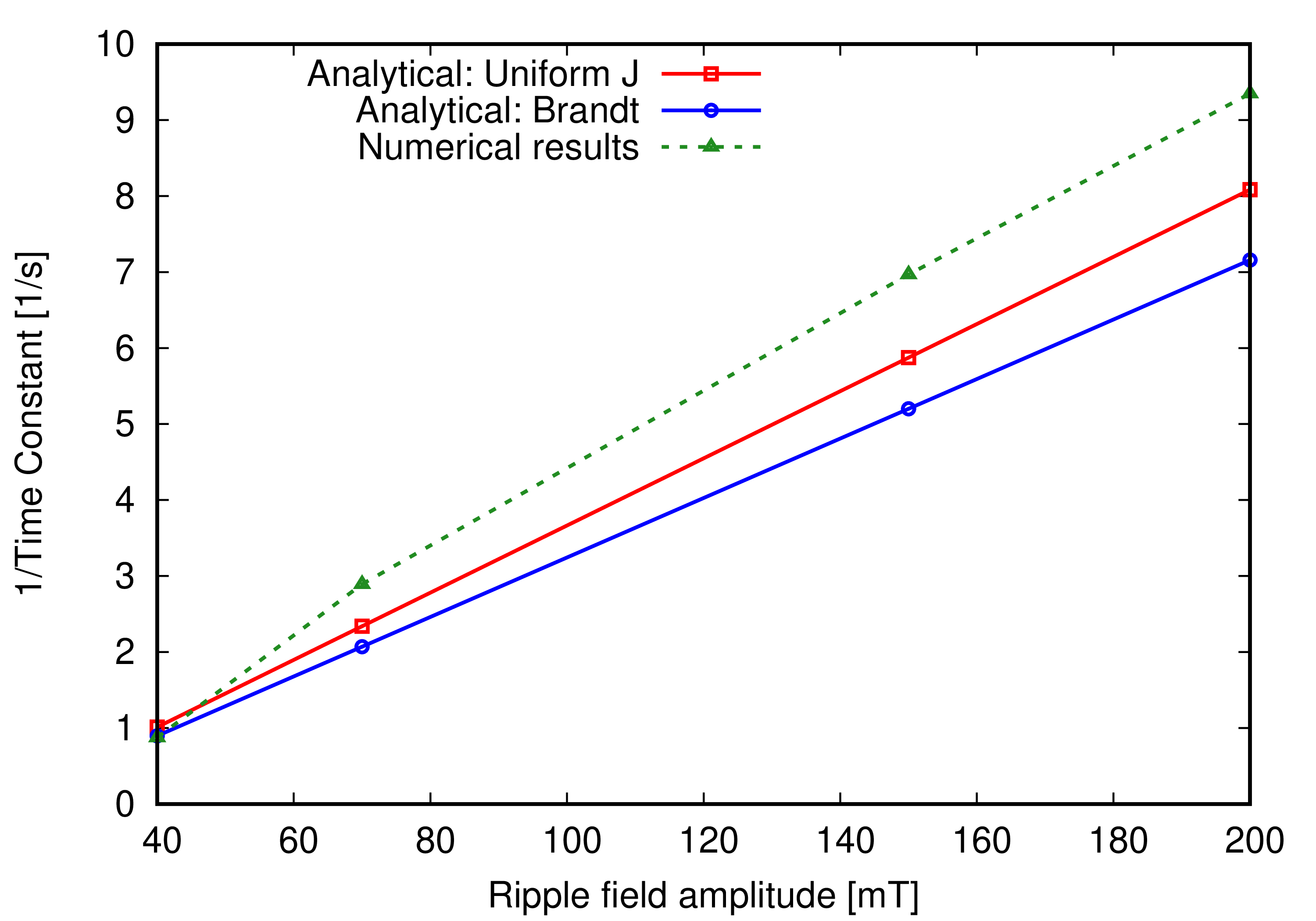}}
\subfigure[][]
{\includegraphics[trim=0 0 0 0,clip,width=8.5 cm]{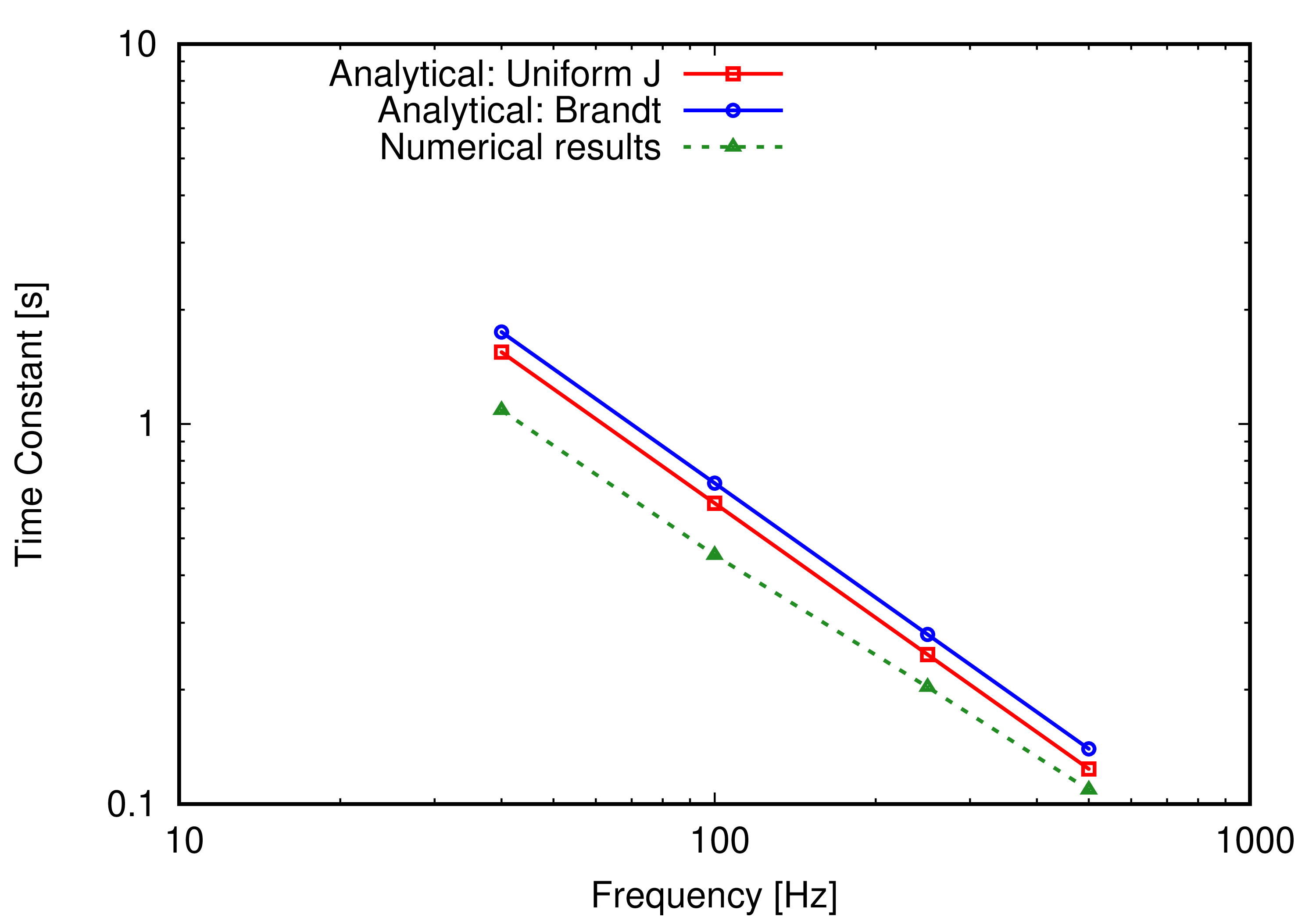}}
\subfigure[][]
{\includegraphics[trim=0 0 0 0,clip,width=8.5 cm]{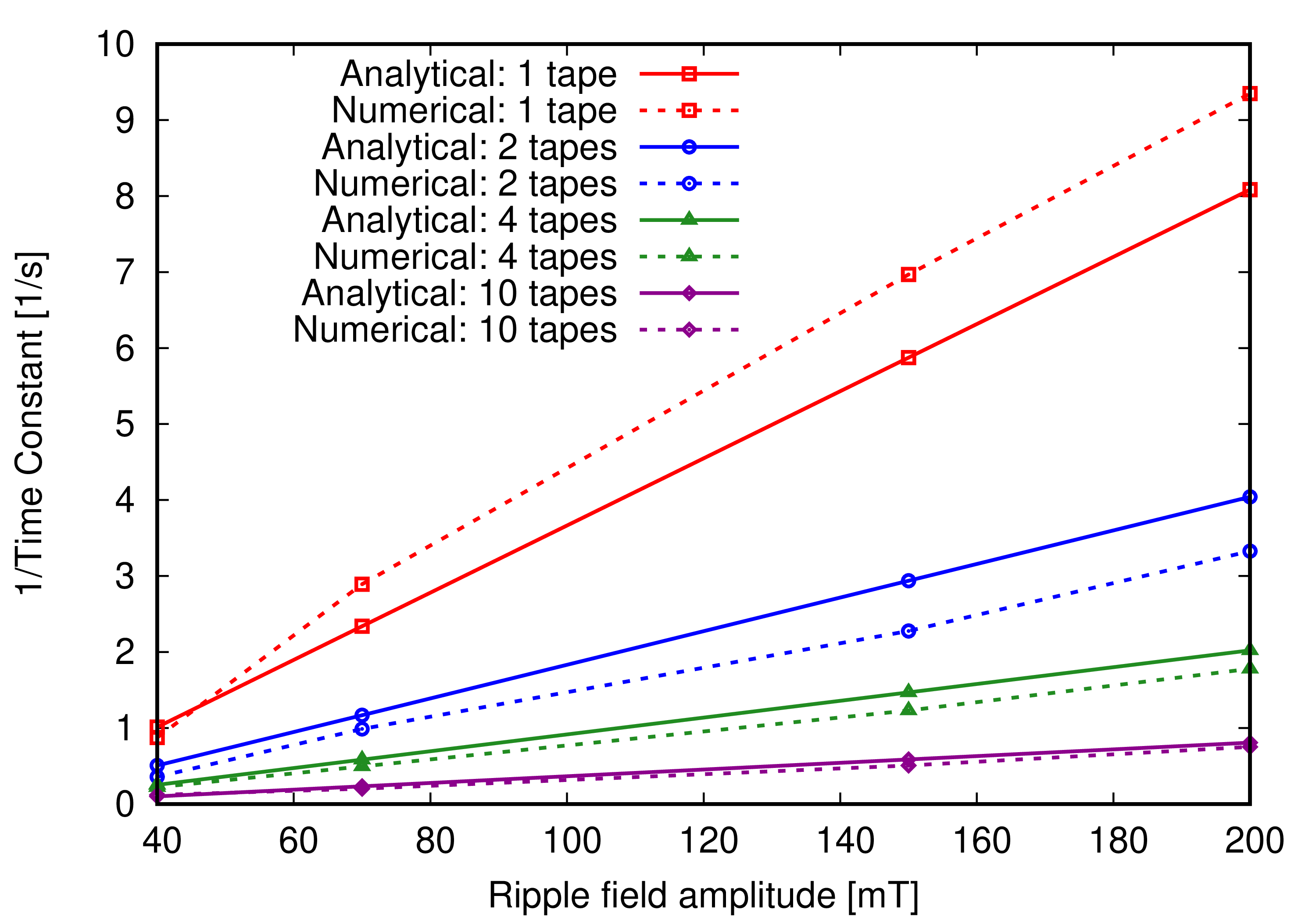}}
\caption{ Time constant dependence on (a) ripple field amplitude, (b) ripple field frequency, and (c) number of tapes, for numerical analysis calculated at 30 cycles. Numerical calculations agree with simplified formulas for single tape (equations (\ref{TC2}) and (\ref{TC_Brandt})), and number of tapes (equation \ref{TC3}). The numerical results get closer to analytical results when calculated for higher $n$ values and higher number of cycles. } 
\label{CTc}
\end{figure}

The numerical results for time constants are also compared with the analytical formulas derived in section \ref{TC_Formulas}. Comparing numerical results for single tape with time constants calculated from equations (\ref{TC2}) and (\ref{TC_Brandt}), we find that the numerical values are very close to the analytical results. These results are supposed to get closer to each other when calculated for higher number of cycles, being the numerical results under-estimated. Also, the analytical results are found using the Critical State Model, so with higher $n$ values for the $E-J$ Power Law, the numerical results will get closer to the analytical ones (Figure \ref{CTc}). Numerical calculations also agree for higher number of tapes (Figure \ref{CTc}), {validating} equation (\ref{TC3}) for its direct use in quick approximation of time constants for a stack.

\section{Conclusion}

Cross field demagnetization is a major issue for HTS motors and its detailed analysis can be done with the use of time constants. 2D MEMEP model is used for this analysis, and it is shown that the trapped field results are {the} same for both 2D and 3D models for long samples. This model is relatively fast and promising for design. From numerical modeling for constant $J_c$, it is observed that the time constant for a HTS stack increases with tape thickness, tape width, and number of tapes in a stack, and decreases with ripple field amplitude and frequency.

The time constant formulas for single tape and stack of tapes derived in the paper are validated by the numerical results, and thus can be used for quick approximation by engineers directly. Equations (\ref{TC2}) and (\ref{TC_Brandt}) both give very similar results. Our formula (equation (\ref{TC2})) predicts a bit lower $\tau$, and hence it is more pessimistic, which is practical for engineering applications. The formula for thick stacks could be used for stacks in a motor environment. {Apart from the predicting power of these formulas, their relatively simple physical background enable researchers to understand the cause of the observed dependencies. For instance, we found that the observed proportional increase in the time constant for thin stacks of tapes is due to the increase of self-inductance of the main magnetization current, rather than the increase in the stack trapped field. Actually, the parallel penetration field of one tape is more relevant than the stack trapped field.}

{This article will ease the design studies of researchers regarding superconducting applications with stacks of tapes. This work also evidences the need of high meshes for reasonably accurate modeling of time constants.} {Future work will be directed to develop faster methods to model up to 100 tapes for millions of cycles.} 


\section*{{Acknowledgements}}

{The authors acknowledge the financial support by the European Union's Horizon 2020 research innovation program under grant agreement No 7231119 (ASuMED consortium), as well as from the Grant Agency of the Ministry of Education of the Slovak Republic and the Slovak Academy of Sciences (VEGA) under contract No. 2/0097/18.}


\bibliographystyle{unsrt}	


\end{document}